\documentclass[11pt]{article}

\usepackage{graphicx} 
\usepackage{subfigure} 

\begin{document}         
% Start your text

\title{\bf A Model Based Approach to Reachability Routing}

\author{ 
Leland~Smith,~Muthukumar~Thirunavukkarasu,~Srinidhi~Varadarajan,\\~and~Naren~Ramakrishnan\\
Department of Computer Science\\
Virginia Tech, Blacksburg, VA 24061\\
Email: \{lbsmith~\vline~srinidhi~\vline~naren\}@cs.vt.edu
}

\date{}

\maketitle

\begin{abstract}
Current directions in network routing research have not kept pace with the
latest developments in network architectures, such as peer-to-peer networks,
sensor networks, ad-hoc wireless networks, and overlay networks.  A common
characteristic among all of these new technologies is the presence of highly
dynamic network topologies.  Currently deployed single-path routing protocols
cannot adequately cope with this dynamism, and existing multi-path algorithms
make trade-offs which lead to less than optimal performance on these networks.
This drives the need for routing protocols designed with the unique
characteristics of these networks in mind.

In this paper we propose the notion of reachability routing as a solution to the
challenges posed by routing on such dynamic networks.  In particular, our
formulation of reachability routing provides cost-sensitive multi-path
forwarding along with loop avoidance within the confines of the Internet
Protocol (IP) architecture. This is achieved through the application of
reinforcement learning within a probabilistic routing framework. Following an
explanation of our design decisions and a description of the algorithm, we
provide an evaluation of the performance of the algorithm on a variety of
network topologies.  The results show consistently superior performance compared
to other reinforcement learning based routing algorithms.
\end{abstract}

\section{Introduction}
The next generation of network technologies such as sensor
networks, peer-to-peer networks, ad-hoc wireless networks, and overlay networks
present challenges that have previously not been witnessed in the Internet
infrastructure.  These networks operate on large topologies which are highly
dynamic in terms of changes in cost and connectivity. In these contexts,
single-path routing protocols, the mainstay on current network topologies,
suffer from either route flap or temporary loss of connectivity when the primary
path fails. In addition, these protocols do not make effective use of the graph
connectivity between a sender and receiver in order to improve performance. 
Effectively, addressing these unique requirements demands routing protocols 
that can address a number of novel performance metrics.
%<need segway line> How did we get here?

% Some historical perspective is in order. 
%(as an aside, pure source routing was tried in the early Internet and, while
%still prevalent in some data center networks, is not scalable for graphs of
%arbitrary diameter). 
Historically, routing algorithms evolved from networks where the only
parameters available for making routing decisions were source and destination
addresses.
\footnote{While source routing is still prevalent in some data center networks, and was
present in the early Internet, it is not scalable for graphs of arbitrary
diameter.} These parameters by themselves do not
have sufficient discriminative capability to avoid loops. Hence optimality
criteria were added to the routing formulation to eliminate loops leading to
single-path routing, which no longer meets the needs of
the next generation of network technologies.

To address the needs of emerging network domains, in this paper we attempt to
build a routing protocol with the following characteristics.  First, the routing
protocol should be capable of converging to a solution even in highly dynamic
environments merely with local information i.e., the protocol does not require
any global knowledge of the topology.  Second, to maximize the bandwidth (and
connectivity) between any pair of nodes, the routing protocol should route along
multiple paths between them.  Third, the routing protocol should route as
efficiently as possible by selecting routes in inverse proportion to their
expected path cost.  Fourth, the protocol should avoid loops as much as
possible and guarantee not to get stuck in loops -- the emphasis is on loop
avoidance rather than loop elimination.  Finally, to be
of maximum practical value, the protocol should work within the confines imposed
by the Internet Protocol (IP) specification, including its header fields which only permit a
source and destination.  As mentioned before, source routing has been tried in IP
networks, but was discarded due to security issues, the lack of space in the IP
header to support full source routing for all nodes, as well as its lack of
scalability in large networks.

Note that these requirements place conflicting demands on routing protocol
design.  Different algorithms make differing trade-offs in this multi-constraint
space.  For instance, distance vector and link state algorithms achieve loop
elimination but are restricted to optimality-based single path routing.  MOSPF
achieves loop-free multi-path routing, only in the restricted case of paths with
identical costs. Hot potato routing achieves true multi-path routing but pays no 
attention to either loops or the `quality' of its paths. The
MPATH~\cite{mpath} algorithm and several of its variants achieve cost-sensitive
loop-free multi-path routing, at the expense of routing table storage overhead
proportional to the number of paths (which can be combinatorial). The
theoretically best, although practically naive, solution would be 
all-sources, all-paths routing.  This achieves the goal of correctness, however 
building and maintaining a complete and correct table of the entire network 
would be impractical for networks of any non-trivial size.  

% (because the Internet infrastructure is too set in its ways). 
While it is still true that source and destination are the only parameters
available for routing on the Internet infrastructure, there is a degree of
freedom thus far unexplored by routing algorithms.  Single-path {\em
deterministic} routing algorithms are driven by a need to achieve loop
elimination at any cost due to the disastrous effects of routing loops in such
algorithms.  However, for a {\em probabilistic} routing algorithm, this does not
necessarily have to be the case.  Therefore, if we relax the requirement for
loop elimination and instead seek to achieve loop avoidance by guaranteeing to
exit loops once they are entered, we are given greater flexibility in laying out
optimization constraints. For this reason, we have chosen to take a
probabilistic approach and to sacrifice loop elimination in favor of loop
avoidance.

The undue emphasis on optimality thus far has created algorithms that aggressively
eliminate loops. This has led to implementations that are intolerant of loops.
On the other hand, the ability to tolerate loops opens up new exploration strategies
for true cost-sensitive multi-path routing that work under the constraints
presented above. We therefore begin with the terminal perspective of {\em reachability
routing}, where the goal is merely to reach a destination. Hot potato routing can be
viewed as a limiting example of reachability routing but we clearly want to do
better. From this perspective, we are in the unique position of being able to explore 
the trade-off between eliminating loops and improving efficiency of selecting paths.

Our specific formulation of reachability routing is probabilistic, multi-path, and
cost-sensitive by efficiently distributing traffic among all paths leading to a
destination.  This type of routing can be viewed as solving an optimization
problem which maximizes the number of paths between two nodes by discovering all
the paths, and then derives the probability to route on a given path by
assessing the path costs leading to the destination.  

In particular, we study reachability routing through the lens of reinforcement
learning, which provides a mathematical framework for describing and solving
sequential Markov decision problems (MDPs). The states are the nodes, the
actions are the choice of outgoing links, and rewards correspond to
path costs associated with the state transitions. A value function imposed on
the MDP (e.g., discounted sum of rewards along a path) essentially leads to
an optimization problem, whose solution is a policy for routing. Intrinsically,
this is what all routing algorithms based on dynamic programming do. However,
single-path routing algorithms learn the best deterministic policy that 
solves the MDP. In this paper, the routing algorithm learns stochastic policies 
that achieve cost-sensitive multi-path routing.

%%Additionally, the process of routing is not only a sequential decision making
%process, but can be considered to be Markovian, as the decisions a router 
%makes on a packet are only a factor of the data contained within the packet, 
%such as destination and arrival port, rather than the packet's previous history.  
%
%Also, due
%to the highly dynamic nature of routing, routing can be considered Markovian,
%because previous state presents an outdated view of the network, and therefore
%is not much use to the decision being made.
%
%In the case of routing protocols, what is 'learned' is a policy that optimizes 
%a value function indicating the type of routing protocol being described.  
%As such, traditional shortest path routing can be viewed as a solution to one 
%type of reinforcement learning problem with a particular value function.  
%Alternatively, reachability routing is the result of solving for another value 
%function.  

Our previous work~\cite{srinidhi03} has indicated that such an approach achieves
true multi-path routing, with traffic distributed among the multiple paths in
inverse proportion to their costs.  In addition, in order for our reachability
routing protocol to be of practical use, we are guiding our design decisions by
the requirement that the protocol work within the confines imposed by the
currently deployed Internet Protocol (IP) architecture.

While multi-path routing is not new, we believe that our notion of reachability
routing represents a promising new direction in the field.  Applying
reinforcement learning in this way is a powerful tool enabling reachability
routing to optimize overall network throughput, while at the same time
providing built-in fault tolerance and path redundancy.  Additional
applications of reinforcement learning within this domain hold the potential to
further optimize routing behavior by adaptively refining the performance
parameters of the algorithm in response to changes in the network topology.

The remainder of this paper is organized as follows: Section II provides an overview
of reinforcement learning, its applicability to network routing, and significant
previous work done on the topic. In Section III we introduce a new model-based
routing algorithm based on RL and describe its implementation in Section IV.
Section V presents evaluation results and Section VI concludes with a summary of
our contributions and directions for future research in the area.

\section{Ants and Reinforcement Learning}
Reinforcement learning~\cite{littman96}~\cite{sutton98} is the process of an agent learning to
behave optimally, over time, as a result of trial-and-error interacting within a dynamic
environment. Reinforcement learning problems are organized in terms of
discrete episodes, which, for the purposes of packet routing, consist
of a packet finding its way from an originating source to its intended destination. 
Routing table probabilities are initialized to small random values, thus enabling 
them to begin routing immediately except that most of the routing decisions will 
not be optimal or even desirable. To improve the quality of the routing decision, 
a router can `try out' different links to see if they produce good routes, a mode of
operation called {\em exploration}. Information learned during exploration can
be used to drive future routing decisions. Such a mode is called {\em
exploitation}. Both exploration and exploitation are necessary for effective
routing.

Our RL routing algorithm is a form of ant-colony optimization~\cite{dorigo99}, in which messages
called {\em ants} are used to explore the network and provide reinforcements for
future packet routing. The ants transiting the network provide intermediate
routers with a sense of the reachability and relative cost of reaching the node
which the ant originated from.  In order to overcome the problems of selective path
reinforcement, which deterministically converge to shortest paths, our model
separates the data collection aspects of the algorithm from the packet routing
functionality, as was proposed by Subramanian~et~al.~\cite{subramanian97}.  In
our model the ants only perform the role of gathering information about the network,
which is then used to guide packet routing decisions. 

Three parameters must be considered when applying ants in a routing framework:
the rate of generation of ants, the choice of their destinations, and the
routing policy used for ants.  RL algorithms perform iterative stochastic
approximations of an optimal solution, so the rate of ant generation directly
affects their convergence properties, shown by Di~Caro~et~al. in AntNet~\cite{dicaro98}.  From a practical
perspective in multi-path routing, we would like to choose destinations for the
ants that will provide the most useful reinforcement updates; hence a uniform
distribution policy assures good exploration. Finally, the policy used to route
ants affects the paths that are selectively reinforced by the RL algorithm. As
our goal is to discover all possible paths, the policy used to route ants should
be independent of that of the data traffic. If we do not separate the policies,
then we would end up with the same problem of selective reinforcement as found
in the Q-routing~\cite{subramanian97} algorithm.

In the context of reinforcement learning using ants, effective credit assignment
strategies rely on the expressiveness of the information carried by the ants.
The central idea behind credit assignment is to determine the relative quality
of a route and apportioning blame. In the case of routing, credit assignment
creates a push-pull effect. Since the link probabilities have to sum to one,
positively reinforcing a link (push) results in negative reinforcements (pull)
for other links.

In the simplest form of credit assignment, called backward learning, ants carry
information about the ingress router and path cost as determined by the
network's cost metrics. At the destination, this information can be used to
derive reinforcement for the link along which the ant
arrived~\cite{subramanian97}. Another strategy, known as forward learning, is to
reinforce the link in the forward direction by sending an ant to a destination
and bouncing it back to the source~\cite{dicaro98}. Subramanian et
al.~\cite{subramanian97} adapt the former approach. Ants proceed from randomly
chosen sources to destinations independent of the data traffic.  Each ant
contains the source where it was released, its intended destination, and the
cost $c$ experienced thus far. Upon receiving an ant, a router updates its
probability to the ant source (not the destination), along the interface by
which the ant arrived.  This is a form of backward learning and is a trick to
minimize ant traffic.

Specifically, when an ant from source $s$ to destination $d$ arrives along
interface $i_k$ to router $r$, $r$ first updates $c$ (the cost accumulated by the
ant thus far) to include the cost of traveling interface $i_k$ in reverse. $r$
then updates its entry for $s$ by slightly nudging the probability up for
interface $i_k$ (and correspondingly decreasing the probabilities for other
interfaces). The amount of the nudge is a function of the cost $c$ accumulated
by the ant. It then routes the ant to its desired destination $d$. In
particular, the probability $p_k$ for interface $i_k$ is updated as:
\[
p_k = \frac{p_k + \Delta p}{1 + \Delta p}, 
p_j = \frac{p_j}{1 + \Delta p}, 
\]
\[
1 \le j \le n, j \ne k
\]
where $\Delta p = \frac{\lambda}{f(c)}, \lambda > 0$ and $f(c)$ is a
non-decreasing function of $c$.

Two types of ants, {\em regular ants} and {\em uniform ants}, are supported to
handle the routing aspect of the algorithm. Regular ants are forwarded
probabilistically according to the routing tables, which ensure that the routing
tables converge deterministically to the shortest paths in the network. Regular
ants treat the probabilities in the routing tables as merely an intermediate
stage towards learning a deterministic routing table. They are good exploiters
and are beneficial for convergence in static environments. With uniform ants,
the ant forwarding probability follows a uniform distribution, wherein all links
have equal probability of being chosen. This ensures a continued mode of
exploration and helps keep track of dynamic environments. In such a case, the
routing tables do not converge to a deterministic answer; rather, the
probabilities are partitioned according to the costs. The constant state of
exploration maintained by the uniform ants ensures a true multi-path forwarding
capability.

\section{Motivation}
Our primary design objective is to achieve cost-sensitive multi-path forwarding, 
while at the same time eliminating the entry of loops as much as possible. We
have made a series of improvements to the uniform ants algorithm proposed by
Subramanian~et~al.~\cite{subramanian97}, culminating in a novel model-based
routing algorithm.

\begin{figure*}
\centering
\includegraphics[scale=0.6]{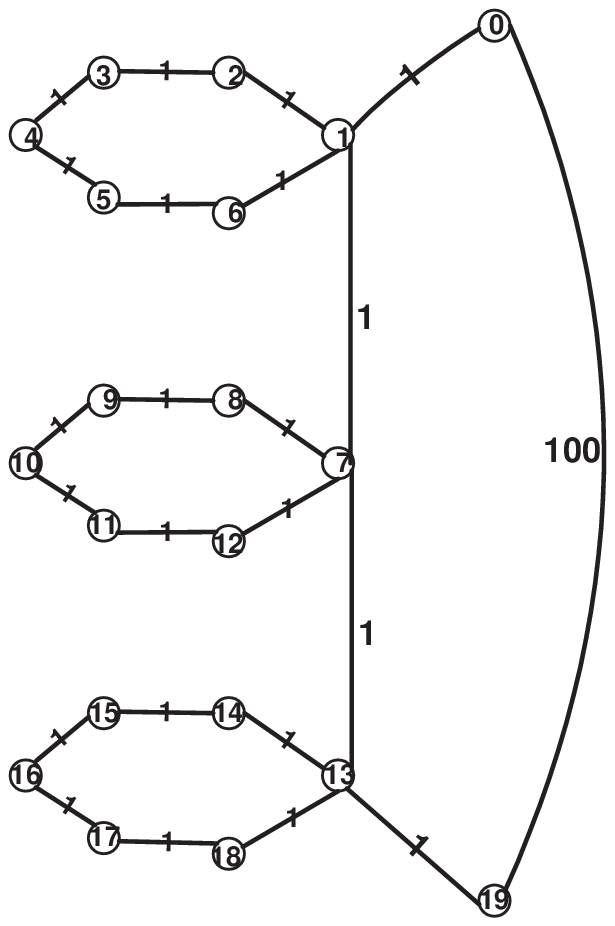}
\vline
\includegraphics[scale=0.6]{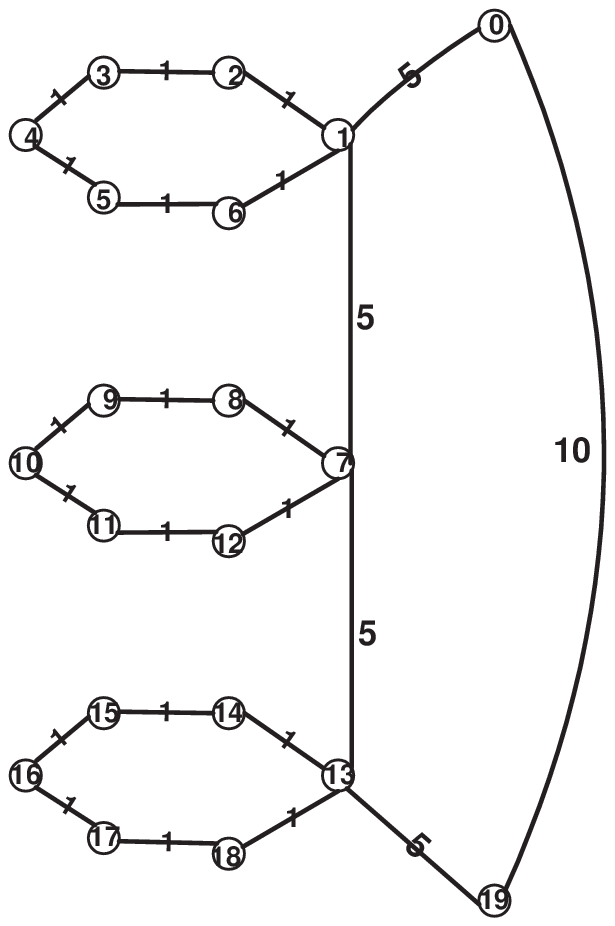}
\vline
\includegraphics[scale=0.6]{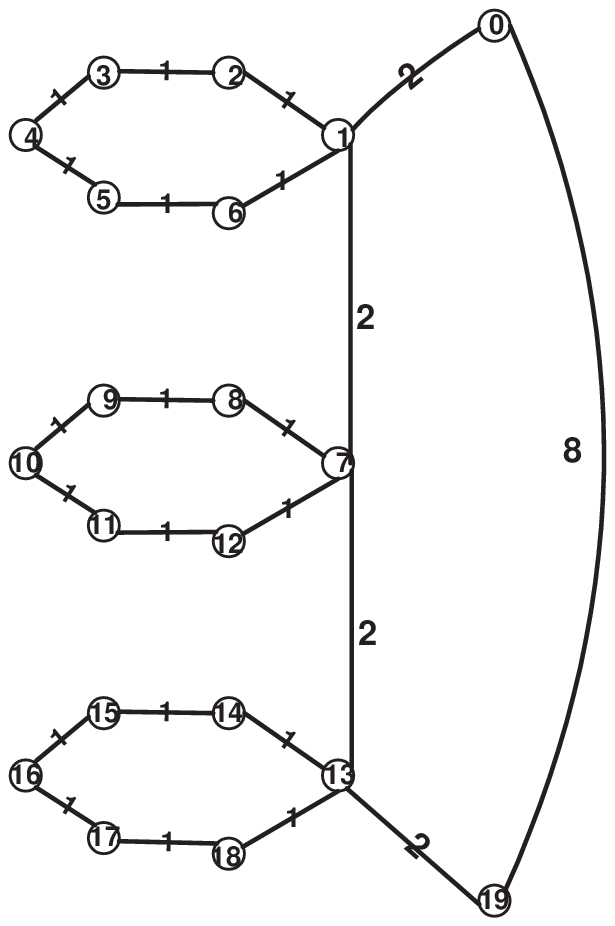}
\caption{Velcro topologies with different cost ratios.}
\label{velcro_topo}
\end{figure*}

Let us begin by observing that uniform ants are natural multi-path routers;
according to Proposition 2 in Subramanian~et~al.~\cite{subramanian97}, the
probability of choosing an interface is aligned in inverse proportion to cost
ratios. The reader might be tempted to conclude that uniform ants inherently support
reachability routing; however consider the three velcro topologies of
Figure~\ref{velcro_topo}.  These topologies have the same underlying graph
structure but differ in the costs associated with the main branch paths (the
direct path from 0 to 19, and the path through nodes 1, 7, and 13).  

Uniform ants explore all available interfaces with equal probability; while this
makes them naturally suitable for multi-path routing, it also creates a tendency
to reinforce paths that have the least amount of decision making. To see why,
recall that the goodness of an interface is inversely proportional to a
non-decreasing function of the cost of the path along that interface. The cost
is not simply the cost of the shortest path along the interface, but is itself
assessed by the ants during their exploration; hence the routing probability for
choosing a particular interface is implicitly dependent on the number of ways in
which a costly path can be encountered along the considered interface.  The
presence of loops along an interface means that there are greater opportunities
for costly paths to be encountered (causing the interface to be reinforced
negatively) or for the ants to loop back to their source (causing their
absorption, and again, no positive reinforcement along the interface). 

The basic problem can be summarized by saying that ``interfaces that provide an
inordinate number of options involving loops will not be reinforced, even if
there exists high-quality loop-free sub-paths along those interfaces.''
Mathematically, this is a race between the negative reinforcements due to many
loops (and hence absorptions), and positive reinforcements due to one (or few)
short or cheap paths. As a result, the interface with the fewer possibilities
for decision making wins, irrespective of the path cost. Hence in the topologies
shown in Figure~\ref{velcro_topo}, uniform ants will reinforce along: the
costliest path (left), among one of many cheapest paths (center) and the
cheapest path (right).  Notice that using regular ants to prevent this incessant
multiplication of probabilities is not acceptable, as we will be giving up the
multi-path forwarding capability of uniform ants.

Ideally, we want our ants to have selective amnesia, behaving as uniform ants
when it is important to have multipath forwarding and morphing into regular ants
when we do not want loops overshadowing the existence of a cheap, loop-free
path. We present a model-based approach that achieves this effect by maintaining
a statistics table independent of the routing table. The basic idea is to make
routers recognize that they constitute the fulcrum of a loop with respect to a
larger path context. 

For instance, in Figure~\ref{velcro_topo}, nodes 1, 7, and
13 form fulcrums of loops, which should not play a role in multi-path forwarding
from, say, node 0 to node 19. The statistics table maintains, for each router
(node) and destination, the number of ants generated by it and the number that returned
without reaching its intended destination. Using these statistics, for instance,
node 1 can reason that all ants meant for destination 19 returned to it, when
sent along the interface leading to node 2. This information can be used to
reduce the scope of multi-path forwarding, on a per-destination basis.  The
statistics table serves as a discriminant function for the choices indicated by
the routing table, while the routing table reflects the reinforcement provided
by the uniform ants.

\section{Protocol Model}
\subsection{Ant Structure}
Ants are small packets used to explore and gather information about
the network. Periodically each source node $s$ generates, to every other destination
$d$, ants of the form $[s, d, c, o_i]$, where $c$ is the cost associated with
the ant and $o_i$ is the outgoing interface from the source router.  When the
ants are created the cost $c$ is initialized to $0$. All the intermediate
routers along the path from the source to destination increment the cost $c$ to
reflect the cost in reverse (when a message traverses a link from node $a$ to
node $b$, $c$ is incremented by the cost of the link from $b$ to $a$). When the
ant reaches the destination $d$, the cost $c$ is the end-to-end cost of sending
a message from source $s$ to destination $d$. Note the intermediate nodes along
the path do not update $o_i$.

\subsection{Routing Table Structure}
The routing table at each node is a two-dimensional array of the probabilities
of using various interfaces to reach destinations.  $RoutingTable_i[j][k]$, maintained at
node $i$, is the probability with which the interface $k$ of node $i$ is chosen
to reach destination $j$. Initially the probabilities for all destinations are
distributed equally across all the interfaces. This is in-line with the
destructive property of RL routing algorithms in which all interfaces are
``innocent until proven guilty.''

\subsection{Statistics Table Structure}
The statistics table is also a two dimensional structure like the routing table,
except each node has two statistics tables.  $SentStatTable_i[j][k]$
corresponds to the number of ants sent along interface $k$ to destination $j$
originating from node $i$, and $ReturnedStatTable_i[j][k]$ is the number of ants
sent along the interface $k$ to destination $j$ which returned to their source
$i$.

The ant statistics are maintained only at the source node, and not at the
intermediate nodes, to allow for scalability of the algorithm.  If every intermediate 
node $n$ along the path of an
ant from source $i$ to destination $j$ increments its statistics table
$SentStatTable_n[j][m]$ when it forwards the ant along the interface $m$, it
would necessitate the ant to have a provision to save the outgoing interface for
each node along its path, so that the node will be able to identify if the ant
loops back to itself. Accommodating such a structure in large topologies would
result in unbounded growth of the ant's size.  Additionally, the ants are not
forwarded when they reach the destination or the source. By updating the
statistics table only at the source nodes, if the ant doesn't loop back to
itself, the source node can safely assume that it has reached the destination
(Under 100\% reliability conditions that no packets are dropped); whereas the
intermediate nodes would have no way of determining whether the ant reached the
destination successfully, or whether it looped back to the source node itself.

\begin{table}
\caption{Model-based Ant Routing Algorithm}
\label{code}
\centering
\begin{verbatim}
procedure Main
  begin:
    Uncontrolled Exploration
    Controlled Exploration
  end.

procedure Exploration (Uncontrolled | Controlled)
  begin:
    for every node in the topology
    begin:
      GenerateAnt; /* Periodically Generate Ant */
      SelectInterface (Uncontrolled | Controlled);
      UpdateModel;
      ForwardAnt;
    end.
  end. /* End of exploration procedure */

procedure ReceiveAnt
  begin:
    if the receiving node is the source of the ant
    begin:
      UpdateModel;
      DestroyAnt;
    end.
    if the receiving node is
      neither the source nor the destination
    begin:
      UpdateRouteTable;
      SelectInterface(Uncontrolled | Controlled)
      ForwardAnt;
    end.
    if the receiving node is the
      intended destination of the ant
    begin:
      UpdateRouteTable;
      DestroyAnt;
    end.
  end. /* End of receive ant procedure */
\end{verbatim}
\end{table}

\subsection{Description of the Algorithm}
An overview of the algorithm is given in Table \ref{code}.  The algorithm
consists of two stages: Uncontrolled Exploration and Controlled Exploration. In
both forms of exploration, each node periodically generates ants destined for
every other node in the topology. The algorithm uses uncontrolled exploration to
collect information about the topology and uses that information to build a
model to control future exploration at the nodes. The information collected
during the controlled exploration is used to update the model as well.  The two
forms of exploration work almost identically except for the SelectInterface
method. The following is a brief description of the various methods used in the
algorithm above.

\subsubsection{GenerateAnt}
This method generates an ant of
the form $[s, d, 0, undefined]$, where $s$ is the source node generating the ant
and $d$ is the intended destination. The initial cost $c$ associated with the
ant is set to $0$. The SelectInterface method determines the output interface,
so at this point, the output interface is undefined immediately after the ant is created.

\subsubsection{SelectInterface}
Due to the probabilistic nature of the routing algorithm, it is essential to 
ensure that the choice of the destination node for each ant at each node is 
uniformly distributed, so that the number of ants generated to the various 
destinations is nearly equal. This method differentiates between the two forms 
of exploration mentioned above, however both forms choose the output interface
uniformly, although the valid interfaces for Controlled Exploration are slightly 
constrained for optimization.

\begin{itemize} 
\item{\bf Uncontrolled Exploration: } Here the choice of the outgoing interface
at each node along the path from the source to destination is unbiased, i.e.
every interface at that node has equal probability of being chosen as the
outgoing interface.  The node generating the ant chooses one interface from its
interfaces and forwards the ant along that interface. If an intermediate node
(not the intended destination node) receives an ant along interface $A$ and
has interfaces other than $A$, it forwards the ant on some interface other than
$A$.  If it does not have any other interface then it sends-back along the
interface $A$ itself.

\item{\bf Controlled Exploration: }Here the choice of outgoing interface is
controlled by a variable called the threshold factor ($\tau$) ranging from $0$ to
$1$. The threshold factor not only affects the multipath capabilities of the routing
algorithm, but also its loop-free capabilities and its correctness with respect to 
the routing of packets (measured by the percentage of packets successfully reaching 
their intended destinations). 
\end{itemize}

Formally, the threshold factor works in the following manner: When a node $i$ 
(source or intermediate) needs to forward an ant intended for destination $j$, 
finds the ratio of $ReturnedStatTable_i[j][k]$ to
$SentStatTable_i[j][k]$ for each of its interfaces $k_1\cdots k_n$. All those
interfaces whose ratios are less than the threshold $\tau$ are eligible for
selection as a forwarding interface.  Then the selection policy is to choose among
the eligible interfaces with equal probability.  Three special cases must be
handled in the case of controlled exploration:
\begin{itemize}
\item{\bf Case 1}
When an ant arrives at a leaf node, i.e. there are no other interfaces other
than the incoming interface, and if it is not the intended destination then the
node sends-back the ant along the same interface.
\item{\bf Case 2}
When all the interfaces at the intermediate node are ineligible, i.e. their
statistic table ratios are above the threshold , then the node sends-back the
ant along the interface it originally received the ant from.
\item{\bf Case 3}
When all the interfaces at the source node are ineligible then the source node
uses the uncontrolled exploration selection policy to break the deadlock. This
case is a very rare occurrence and occurs only when  is set to a very low value.
\end{itemize}

Once the outgoing interface is selected the next step is to forward the ant
along the chosen interface (ForwardAnt).  In the case of source node, before
calling the ForwardAnt, UpdateModel is called to update the statistics table.

\subsubsection{UpdateModel}
This method updates the statistics tables when an ant is generated or loops back
to its source.  The correctness and currency of the statistics tables are vital
to the performance of the router.  When the node generates the ant $[i, j, c, k]$, 
it increments its statistic table entry $SentStatTable_i[j][k]$ by $1$ to indicate 
that interface $k$ was chosen by $i$ to forward the ant intended for destination $j$.
Also, when an ant $[i, j, c, k]$ loops back to the source node, the statistic table
entry $ReturnedStatTable_i[j][k]$ is incremented by $1$ to indicate that the
choice of interface $k$ to route the ant intended to destination $j$ resulted in
a loop.  This can be considered a negative reinforcement in the behavior of the
router.

\subsubsection{ForwardAnt}
This method is used to forward the ants from the current node to the next node 
along the interface chosen by the SelectInterface method.

\subsubsection{DestroyAnt}
When the ant reaches the intended destination or loops back to its source
itself, the ant is not forwarded further and the node absorbs the ant.

\subsubsection{UpdateRouteTable}
When any node $t$ (intermediate or the intended destination) other than the
source node, receives an ant $[i, j, c, k]$ on interface $l$ from node $y$, it
updates the cost $c$ by adding the cost of traversing the interface $l$ in
reverse, and then updates its routing table entries for node $i$ as follows:
\[
rt[i][l] = \frac{rt[i][l] + \Delta p}{1 + \Delta p}, 
rt[i][m] = \frac{rt[i][m]}{1 + \Delta p}
\]
\[
1 \le m \le n, l \ne m
\]
where $\Delta p = \frac{\lambda}{f(c)}$, $\lambda > 0$ and $f(c)$ is a
non-decreasing function of $c$.

\subsection{Qualitative Characteristics}
The model-based routing algorithm presented above discards all {\em useless
loops}, in which all traffic exiting the loop must exit at the same point which
it entered, such as the fulcrum points in the velcro topologies shown in
Figure~\ref{velcro_topo}.  For instance, in these velcro topologies,
when node 1 sends out a packet intended for a destination other than those nodes
in the loop pivoted at 1, either on the interface leading to node 2 or node 6,
the result will be the packet returning to node 1. From the statistics table,
node 1 will learn that those interfaces are useless for forwarding packets to
certain destinations and hence avoid them in the future. By discarding all the
useless loops, this algorithm overcomes the problem of the uniform ants
algorithm wherein only the path with the least decision-making is reinforced.

The threshold factor $\tau$ influences the reinforcement of the various paths of
a topology. At very high values of $\tau$, the algorithm tends towards behaving
like uniform ants while continuing to avoid all the useless loops. For instance
a $\tau$ value of 1 means that an interface where all but one packet sent on it
looped back may still be selected as an outgoing interface. At the same time
this setting still avoids all the interfaces that lead to useless loops, as all
packets sent along them must have come back to the sender. 

At high $\tau$ values, certain packets may encounter one or more loops along
their path that are unavoidable. At very low values of $\tau$, the nodes have a
limited selection of interfaces to choose from due to the stringent
loop-avoidance criteria, which will affect our goal of multi-path routing, but
will greatly decrease the probability of encountering a loop.  The choice of
$\tau$ factor determines the multipath, correctness, and loop-avoidance
capabilities of our algorithm. The threshold factor can either be set to a fixed
value (for the network, or on a per-router or per-router/destination-pair basis)
or can be adaptively refined to optimize model-based routing for various
criteria.

\section{Evaluation}
\subsection{Experimental Setup}
To measure the performance of our cost-sensitive reachability routing algorithm,
we wrote a discrete event simulator in C to simulate a standard
point-to-point topology based network. The simulated network is modeled as a set
of nodes interconnected over point-to-point links, each with an associated cost. The
discrete event simulator was derived from work done in~\cite{srinidhi00}, and
has been used in several networking courses to model routing algorithms.

The simulator runs at a resolution of 1 $\mu s$ and an integer value defined at
the initialization of the simulation determines the duration of the simulation.
In our case, the simulation runs were set to INTMAX (2147483647 as defined in
$<$ limits.h $>$).  As it is a discrete event simulator, every action takes
place after the expiration of a timer and the simulator is programmed to run in
uncontrolled exploration mode for the first one eighth of the time and in
controlled exploration mode for the remaining time. Each node generated an ant
every 10000 $\mu s$. For the purpose of this paper, we programmed the link layer
of the simulator to be reliable, i.e. it does not introduce any errors or drop
packets.

\subsection{Topologies}
A utility provided along with the simulator~\cite{srinidhi00}, when given the
number of nodes in the network and number of interfaces per node, is able to
generate four different interconnected topologies for the network, namely: tree,
clique (fully connected mesh), arbitrary graph, and loop topologies. The automated 
topology generating utility was used to generate the tree and arbitrary graph 
topologies used in the simulations.

Using the manual topology generator provided along with the simulator, complex
topologies such as the velcro and dumbbell topologies were created. These
topologies have some intrinsic characteristics helpful in demonstrating the
range and effectiveness of our algorithm.

A clique topology generator was written in C, which, when given the number of rows and
columns in the clique, will generate a perfect clique topology wherein all the
interior nodes will be of degree 4 and all the boundary nodes will be of degree
2 or 3.

Finally, BRITE, the Boston university Representative Internet Topology
gEnerator~\cite{brite}, was used to generate large Internet scale topologies. It
provides a wide variety of generation models, as well as the ability to extend 
such a set by combining existing models or adding new ones. We used the
Router Waxman Flat Router-level model, which is governed by a power law, to
generate the topologies. A program in C was written to convert the topology
format generated by BRITE to the format used by our simulator.
Topologies with sizes ranging from 20 to 200 nodes were generated using BRITE.

Our model-based routing algorithm was first validated in~\cite{srinidhi03}, by
examining its performance when applied to routing on synthetic worst-case
scenario topologies, such as velcro topologies.  This previous work also
presented a subtle modification to the algorithm, avoiding sub path reinforcement, 
which results in better performance on certain types of topologies. 

% Second, we quantify the convergence
% of our routing algorithm by measuring the correlation of path costs and hop
% counts between all packets sent to and originating from the nodes under
% consideration. In our case, the nodes under consideration were those with the
% maximum and minimum degree. 

The primary contribution of this work is to study data traffic across the network
based on converged routing tables and introduce a new factor called the
reachability factor ($\phi$) that controls the choice of the outgoing
interfaces. We investigate the effect of the threshold factor ($\tau$) and the
reachability factor on various topologies with the help of an operating curve
aimed at helping network administrators in choosing the ideal threshold and
reachability factors for their networks. We also show that by making the nodes
always choose the interface with the highest probability for the intended
destination, our model-based routing algorithm behaves in the same way as any
other single-path deterministic routing algorithm i.e., it provides loop-free
shortest-paths with guaranteed delivery for all packets. 

Additionally, we show that even though the goal of every multi-path routing
algorithm is to avoid loops, our model-based routing algorithm does not
guarantee a complete elimination of loops.  Nevertheless our algorithm
guarantees that a packet will eventually exit the loop and reach its intended
destination. We study the distribution of loops encountered by packets and show
that a vast majority of packets encounter only a small number of loops, or none
at all.

\subsection{Packet Routing Using Model Based Routing}
In this set of experiments, a new application was written on top of the
simulator to route packets based on the routing table learned by ants exploring
the network. Initially, we ran the model-based routing algorithm on the given
topology to obtain a stabilized routing table. Next, we ran the application with
the routing table and the reachability factor as parameters and collected
various statistics. Below we will discuss in detail the application, the
reachability factor, the statistics collected, and analysis of the statistics
obtained from both model-based and uniform ants routing.

The functioning of the application is similar to the one described earlier,
except that there is no update of the routing table.  The routing table is
pre-initialized to that obtained from the model-based routing simulation and
remains constant throughout. By not updating the routing table based on the
packets arriving at every node we are just exploiting the model and not
exploring the network further.  It should be noted that a real world router
would constantly explore the network with ants, and use the resulting routing
table to route packets simultaneously.  However, to determine the effectiveness
of the underlying algorithm, it is simpler to analyze its performance in a
static network environment.

The reachability factor $\phi$  controls the degree of freedom each node has in
choosing the outgoing interface. At each node the outgoing interfaces are
ordered in descending order of their probabilities for every destination. When a
node $n$ needs to route a packet intended for destination $d$, it picks the top
$\phi$ interfaces for that destination and uses their scaled up probabilities
for selecting the outgoing interface. For a better understanding of the
reachability factor, consider the following example. Say a node $M$ has 4
interfaces $A$, $B$, $C$, and $D$ with associated probabilities $0.4$, $0.2$,
$0.15$, $0.15$ for destination $N$; then a $\phi$ value of $2$ will allow the
node $M$ to choose from interfaces $A$ and $B$ with probabilities $(0.4)/(0.4 +
0.2)$ and $(0.2)/(0.4+0.2)$ respectively i.e. node $M$ will choose interface $A$
66.67\% of the time and interface $B$ 33.33\% of the time to route the packet
intended for destination $N$.

The statistics collected include the number of loops encountered by the packets
along their paths, the number of packets encountering loops, the multipath
capability of the packets, and the percentage of packets successfully reaching
their intended destination. To determine the number of loops encountered by the
packets, each packet has a stack associated with it. Every node, before
forwarding a packet, checks to see if its {\em id} already exists in the stack. If its
{\em id} is present in the stack, it increments the loop counter of the packet by $1$
and pops the contents of the stack up to its {\em id} else pushes its {\em id} onto the
stack and then forwards the packet. At the end of the simulation we have
statistics on the number of packets encountering loops (loop percentage) and the
total number of loops encountered by all the packets. Every packet also has a
multipath flag associated with it that is set if any node along the path taken
by the packet has more than one outgoing interface to choose from. This is used
to determine the percentage of packets that could have potentially taken more
than one path to reach their intended destination (multipath percentage).
Finally, we determine the success percentage as the percentage of packets
successfully reaching their intended destination.

\subsubsection{Reachability factor $\phi = 1$}
In our first set of experiments $\phi$ was set to $1$ so that the nodes always
choose the best outgoing interface (interface with the highest probability) for
each packet. As each packet deterministically chooses the best interface at
every node, the multipath percentage is zero. A $\phi$ value of $1$ also results
in the avoidance of loops and a one hundred percent success percentage as all
the packets reach their intended destination. According to proposition 2 of
Subramanian~et~al.~\cite{subramanian97} the probability of choosing an interface
is inversely proportional to the cost ratios (under the assumption of loop free
paths).  Keep in mind that this proposition applies even for our modified
model-based algorithm as all the avoidable loops are avoided and also we have
shown in~\cite{srinidhi03} that the probabilities are inversely proportional to
the path costs. By choosing the interface with the highest probability, i.e. the
interface that advertised a lower cost path to that destination, at every node
we have achieved deterministic shortest path routing while still using the
underlying probabilistic routing table. 

The following set of simulations were done on 20 to 100 node BRITE topologies
with uniform cost distribution so that with $\phi = 1$ the path taken by all the
packets will not only correspond to the shortest path in terms of cost but also
in terms of the number of hops. By sending packets across the network and
keeping track of their hop count, we ascertained the shortest path length
between every source-destination pair. At the end of the simulations, the
average shortest path length for the topologies were calculated and compared
with the theoretical shortest path lengths. We then attempt to fit this
empirical data onto parametrized formulas.

Below we discuss the derivation of average shortest-path lengths for
exponentially distributed graphs based on~\cite{newman01}.  The Router Waxman
model of BRITE uses an exponentially distributed generation function to create
the topologies.  According to~\cite{newman01}, the generating function $G_0(x)$
should be normalized such that $G_0(1) = 1$.

We use the following generating function for our derivation:
\[
G_0(x) = \frac{1 - e^{-1/\kappa}}{1 - xe^{-1/\kappa}}
\]
According to~\cite{newman01}, the average shortest path length is given
by:
\[
l = \frac{\ln{N / z_1}}{\ln{z_2 / z_1}} + 1
\]
for $N \gg z_1$ and $z_2 \gg z_1$, where $N$ corresponds to the
number of nodes in the topology, and $z_m$ corresponds to the
average number of $m$th-nearest neighbors with $z_1 = G'_0(1)$
and $z_2 = G''_0(1)$. We derived $l$ to be:
\[
l = 1 + \frac{\ln{N} + \ln{e^{1/\kappa}-1}}{\ln{2} - \ln{e^{1/\kappa} - 1}}
\]

From this equation we derived the value of $\kappa$ to be
\[
\kappa = \frac{1}{\ln{\frac{2^{\frac{l-1}{l}}}{N^{\frac{1}{l}}}}}
\]

\begin{figure}
\vspace{0.14in}
\centering
\includegraphics[scale=0.35]{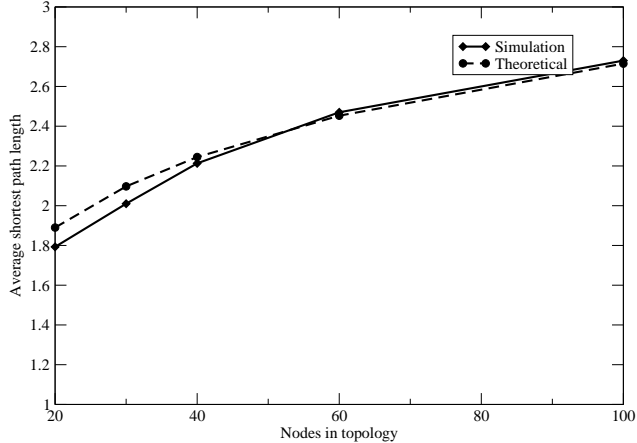}
\vspace{0.14in}
\caption{Least square fit between the theoretical and actual shortest path
lengths.}
\label{shortest_path}
\end{figure}

Based on the above derivations, a least square fit was conducted on the
simulation results, which returns both $\kappa$ and the square of the
correlation coefficient with values ranging of $0$ and $1$, indicating bad or
good fit respectively. In our case, the fit returned a value of $0.986551$,
which indicates that the best fit line summarizes the data very well as shown in
Figure~\ref{shortest_path}.

\subsubsection{Reachability factor $\phi =$ maximum degree}
\begin{figure*}
\centerline{\subfigure[With source absorption, each point is labeled with its
threshold value and success percentage]{\includegraphics[scale=0.35]{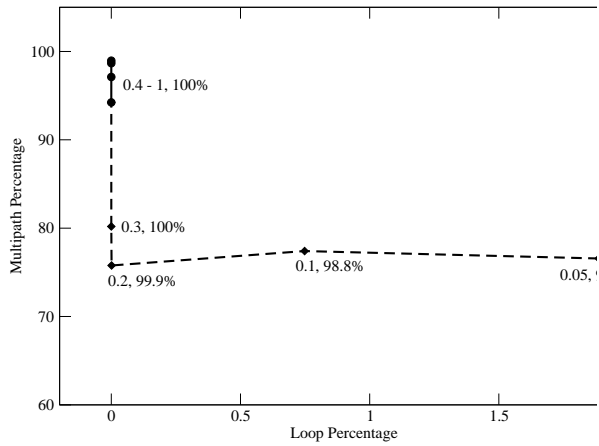}
\label{curve_velcro_abs}}
\hfill
\subfigure[Without source absorption, each point is labeled with its
threshold value]{\includegraphics[scale=0.35]{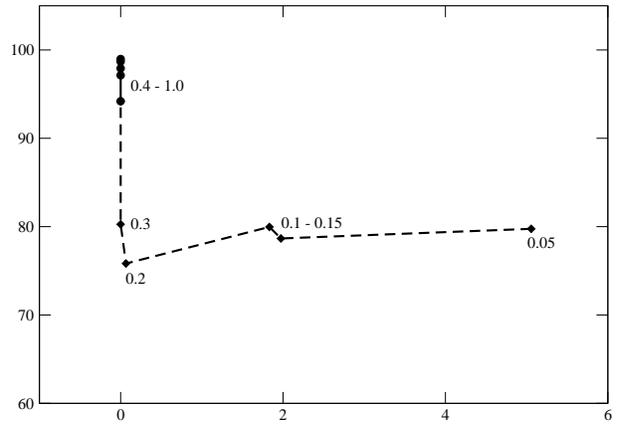}
\label{curve_velcro_noabs}}}
\caption{Operating curve for the velcro topology shown in
Figure~\ref{velcro_topo} right}
\vspace{0.14in}
\label{curve_velcro}
\end{figure*}

\begin{figure*}
\vspace{0.14in}
\centerline{\subfigure[With source absorption, each point is labeled with its
threshold value and success percentage]{\includegraphics[scale=0.35]{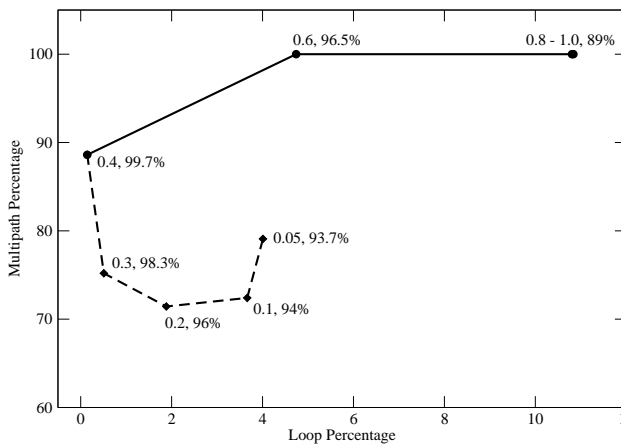}
\label{curve_clique_abs}}
\hfill
\subfigure[Without source absorption, each point is labeled with its
threshold value]{\includegraphics[scale=0.35]{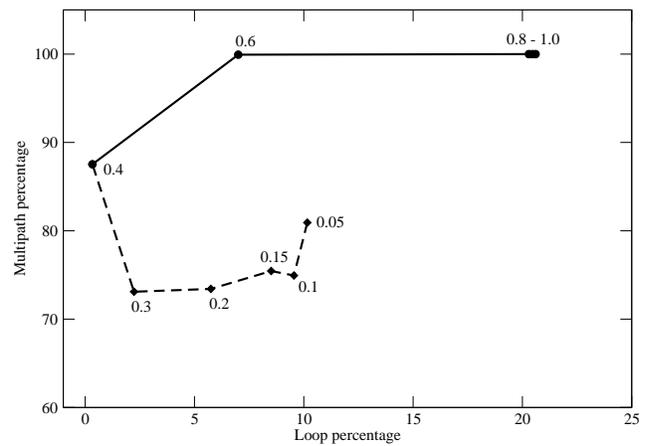}
\label{curve_clique_noabs}}}
\caption{Operating curve for a 8x5 clique topology}
\label{curve_clique}
\end{figure*}

By setting the reachability factor to the maximum degree of the topology, each 
node will be allowed to choose among all its interfaces to be the outgoing
interfaces (based on the probability associated with it for the intended
destination). The simulations were run on the following topologies: 20 to 200
node BRITE topologies, 10x4 \& 8x5 clique topologies and the velcro topologies
described in Figure~\ref{velcro_topo}. {\em Operating curves} of the percentage 
of packets encountering loops were plotted against the percentage of those with 
multipath capabilities for various topologies at different values of the threshold 
factor.  These operating curves are shown in 
Figures~\ref{curve_brite},~\ref{curve_velcro},~and~\ref{curve_clique}.  
Visualizing the performance of the routing algorithm in this way enables us to
compare the effect of the inherent topology and performance parameter settings,
and the interactions between the two.

As opposed to $\phi = 1$, $\phi =$ maximum degree results in multi-path
forwarding of the packets and also some portion of packets entering into
loops. All the packets reached their intended destinations except for those that
looped back to their source resulting in a high success percentage. To overcome
the drop in success percentage, the packets were forwarded even when they looped
back to the source and counting this episode as just another loop encountered
along the path.  

With this modification all the packets successfully reached
their intended destinations but with a linear increase in the percentage of
loops (to account for all those packets that were earlier absorbed by their
source). All packets had a TTL of 255 but none of them were dropped due to
reaching the TTL limit. Below we present the operating curves for various
topologies under both the cases: 1) absorption of packets at their source and 2)
no absorption of packets. 

\subsection{Operating Curve Observations}

\begin{figure*}
\vspace{0.14in}
\centerline{\subfigure[With source absorption, each point is labeled with its
threshold value and success percentage]{\includegraphics[scale=0.35]{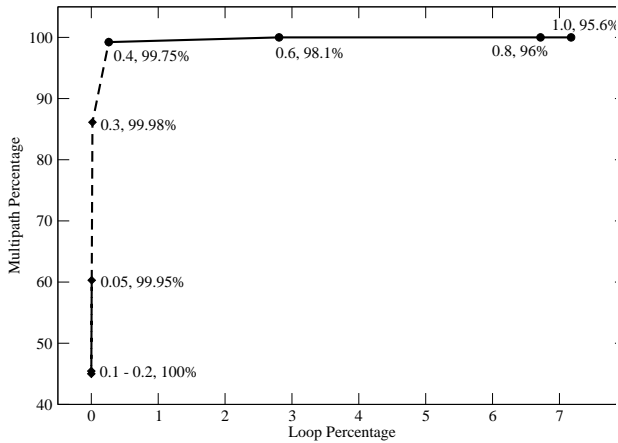}
\label{curve_brite_abs}}
\hfill
\subfigure[Without source absorption, each point is labeled with its
threshold value]{\includegraphics[scale=0.35]{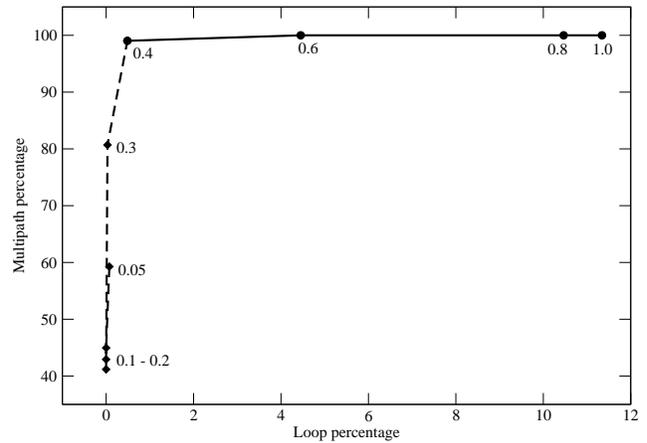}
\label{curve_brite_noabs}}}
\caption{Operating curve for a 40 node BRITE topology}
\vspace{0.14in}
\label{curve_brite}
\end{figure*}

Let us take the operating curve for a random 40 node BRITE topology shown in
Figure~\ref{curve_brite_abs} and study it closely. As the threshold factor
increases, we see that the performance goes from a region with no loops and 45\% multipath to
one with 7\% loops and 100\% multipath. It is heartening to note that the curve first
increases in the direction of accommodating multipath before introducing loops,
rather than the other way around.

Second, notice that different portions of the graph are shaded differently. Each
operating curve is represented by a solid line and a dotted line. These denote
the region where the model is completely in force, and the region where it is
not, respectively. As discussed earlier, at very low threshold factor values,
when all the interfaces at an intermediate node are ineligible, i.e. their
statistic table ratios are above the threshold, then the node sends-back the ant
along the interface it originally received the ant from resulting in an
increased percentage of packets entering into loops. Similarly at very low
values of $\tau$, when all the interfaces at the source node are ineligible,
then the source node uses the uncontrolled exploration selection policy to break
the deadlock. As Figure~\ref{curve_brite_abs} shows, around a threshold value of
$0.4$, the model comes into force in that all routing decisions are based on
learning rather than defaults.  

By comparing Figure~\ref{curve_brite_abs}
with~\ref{curve_brite_noabs} (the latter of which does not have source
absorption), we notice that the difference in the percentage of success of
packets reaching their destination with and without source absorption is
reflected in the difference in percentage of packets encountering loops with and
without source absorption.  Removing source absorption from the simulation
results in a 100\% success rate, but an increase in the percentage of packets
encountering loops, which is an understandable consequence.  However, for a
router using the ant-derived statistic tables to make routing decisions, it is
vital for data to transit the network with the highest success rate, even at the
expense of an increased likelihood of entering a routing loop.

The operating curves for the 40 node BRITE topologies shown in
Figures~\ref{curve_brite_abs} and \ref{curve_brite_noabs}, compared to the
operating curves of BRITE topologies with different numbers of nodes (not shown
here, refer to~\cite{kumar04}) also exhibit another
interesting behavior. As the number of nodes in the topology increases, the
minimum multipath percentage also increases. This is due to the fact that at
very low threshold values, the model-based routing algorithm routes a large
number of packets deterministically in smaller topologies.  The shape of the
operating curve greatly depends on the intrinsic graph theoretic property of the
topologies. The reader can observe from the figures above that each topology
class (BRITE, clique, and velcro) generates its own unique shape of operating
curve.  

% Figure 4.28 and Figure 4.29 have their
% operating curve very similar to the operating curve generated
% by the mesh topologies as the topology in Figure 4.6 can be
% viewed as a triangulated mesh topology.

The reader should also observe that all the operating curves at $\tau = 1$
exhibit the behavior of the uniform ants algorithm~\cite{subramanian97}.  This
is due to the fact that all the interfaces at each node are eligible to be
selected as the outgoing interface for the intended destination which conforms 
to the selection policy of uniform ants algorithm.

The number of unique operating curves is limitless when the various topology
classes are combined in the same network.  The fact that each operating curve
has a unique threshold value that gives the network optimal performance, in
terms of loop avoidance and multipath routing, presents us with the need to
adaptively learn and refine that threshold value for an arbitrary dynamic
network.  This is an area of future research that is necessary before our
multipath routing algorithm can be deployed on actual networks. 

\subsection{Distribution of loop frequency}
\begin{figure}
\vspace{0.14in}
\centering
\includegraphics[scale=0.35]{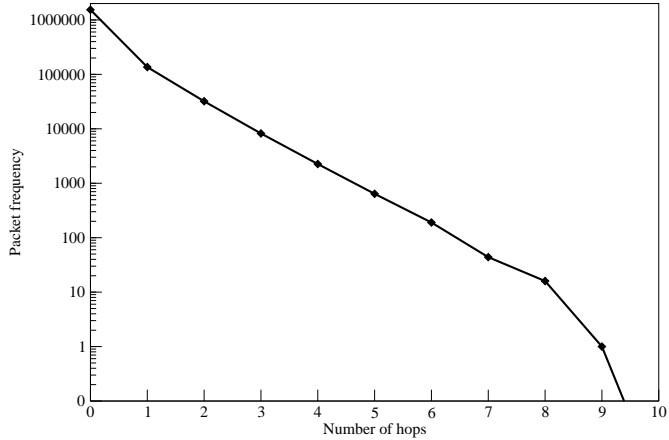}
\vspace{0.14in}
\caption{Distribution of loops encountered by packets in a random 40 node BRITE
topology, note the logarithmic scale of the y-axis.}
\label{loop_freq}
\end{figure}

Finally, we show that even though the presence of loops is unavoidable, the
number of packets that encountered $k$ loops along their paths to their
respective destinations exponentially decays with increase in $k$, i.e. the
majority of the packets encounter between 0 to 2 loops, suggesting a power law.
Figure~\ref{loop_freq} shows the plot between loop distribution and packet
frequency for a 40-node BRITE topology. It should be noted that due to the 
cyclic nature of clique topologies, certain packets in those topologies encounter 
as many as 20 loops before they reach their intended destination.

\subsection{Verification of cost-sensitive routing in BRITE topologies}
The goal of this experiment was to perform a large-scale validation of the
cost-sensitivity properties of our reachability routing algorithm. First, all
the paths taken by various packets at $\phi =$ maximum degree were enumerated.
To achieve this, every packet had a stack associated with it that kept track of
the nodes visited by it en route to its destination. At the destination, the
paths taken by the packets from each source were ranked in increasing order of
their costs. The destination nodes only kept track of unique paths from each
source and also maintained the frequency associated with each path. 

\begin{figure*}
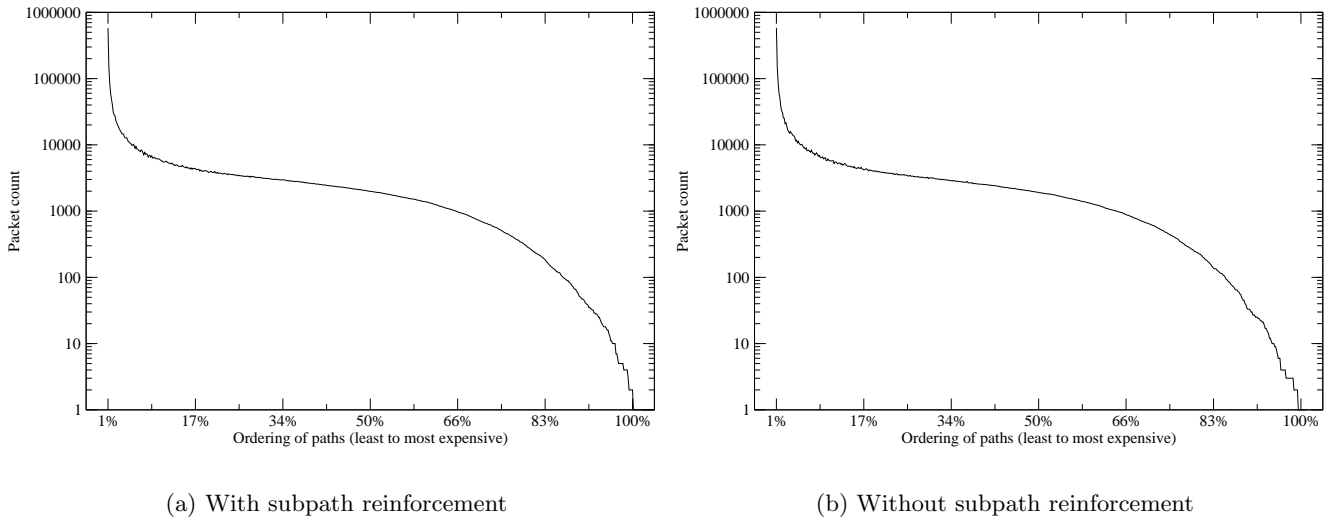

\vspace{0.14in}
\centerline{\subfigure[With subpath reinforcement]{\includegraphics[scale=0.35]{dist_subr_noabs}
\label{traffic_dist_subr}}
\hfill
\subfigure[Without subpath reinforcement]{\includegraphics[scale=0.35]{dist_nosubr_noabs}
\label{traffic_dist_nosubr}}}
\caption{Traffic distribution in a random 60 node BRITE topology without source
absorption, note the logarithmic scale of the y-axis.}
\vspace{0.14in}
\label{traffic_dist}
\end{figure*}

The purpose of this instrumentation was to ensure that the frequency of costs, as measured
through pursued paths, mirrored the distribution of traffic along these paths.
At the end of the simulation, the summation of frequency over the top $[x - 9\%,
x\%]$ of the paths for every source-destination pair was determined, for $x \in
[10, 20 \cdots 100]$.  Figures~\ref{traffic_dist_subr}
and~\ref{traffic_dist_nosubr} show the cost-sensitive routing of our
model-based algorithm and also that sub-path reinforcement has no effect on
BRITE topologies. (The experiments were performed on 60-node BRITE topologies).

\section{Conclusion and Future Work}
In this paper we have presented a new model-based reinforcement learning
algorithm, which achieves true cost-sensitive reachability routing, even in
network topologies that pose problems to both deterministic routing as well as
classical RL formulations.  This algorithm efficiently distributes traffic among
all paths leading to a destination. The evaluation results indicate that our
approach achieves true multi-path routing, with traffic distributed among the
multiple paths in inverse proportion to their costs. By helping maintain the
incremental spirit of current backbone routing algorithms, this approach has the
potential to form the basis of the next generation of routing protocols,
enabling a fluid and robust backbone routing framework. The reader is referred
to~\cite{kumar04} and~\cite{srinidhi03} for background and further experimental
results.

We now present four possible directions for future work.

\begin{itemize}
\item{\bf Adaptive configuration of the threshold factor ($\tau$)}

The threshold factor is currently set to a fixed value for all the nodes in the
topology. From the operating curve, the network administrator determines the
optimal value of $\tau$ at which the routing yields high success and multipath
percentage while keeping the percentage of packets entering into loops low. As
part of the future work, we can determine the $\tau$ value dynamically based on
available information and periodically adjust its value to obtain the optimal
routing requirements. The  value could be dynamically adapted on a per-node
basis or on per-source/destination-pair basis at every node.

\item{\bf Instructive feedback}

Our RL algorithm works primarily using evaluative feedback from neighboring
routers. It would be interesting to extend the framework to accommodate
instructive feedback. But to provide instructive feedback, a router must have
sufficient discriminating capability to perform credit assignment. It is
typically of the case that any resulting instruction will be of the negative
kind i.e., ``for destination $X$, do not use interface $i_y$.'' How such
negative instructions can co-exist with positive reinforcements is an important
research issue, not only for our application domain, but also the larger field
of reinforcement learning.

\item{\bf Modeling topologies with hierarchical addressing}

Currently the algorithm assumes all topologies to be flat such that all nodes in
the topology are numbered from $1$ to $n$. By supporting hierarchical addressing
of the nodes, the model built at every node could be at a sub network basis
instead of being at a per node basis, i.e. a node could collect statistics for a
group of nodes as a single entity and build its model accordingly. Such an
approach encourages problem decomposition and enables scaling up to large
network sizes.

\item{\bf Reverse engineering routing protocols}

The model-based reinforcement learning algorithm presented here promises to
serve as an abstraction of reachability routing algorithms in general. One idea
for further research is to automatically mine the model by analyzing implemented
routing algorithms' behavior, rather than incrementally learning it from
scratch, as we have done here. In other words, we can seek to imitate the
functioning of another algorithm by suitably configuring our model.  This
problem has its roots in inverse reinforcement learning, where we are aiming to
recover an algorithm from observed (optimal) behavior.  The first steps toward
such reverse engineering have been recently taken~\cite{shiraev03}.

\end{itemize}

\end{document}